\documentclass[conference]{IEEEtran}
\usepackage{algorithm, algorithmic}

\usepackage{ifpdf}
\ifCLASSINFOpdf
\else
\fi
\usepackage{url}
\usepackage{amsmath,graphicx}
\usepackage{placeins}
\usepackage{amsmath,epsfig}
\usepackage{epstopdf}
\usepackage{amssymb}

\usepackage{dblfloatfix}
\usepackage{amssymb}
\usepackage{caption}
\usepackage{subfigure}
\usepackage{pstricks}
\usepackage{subfloat}
\begin{document}
%
% paper title
% can use linebreaks \\ within to get better formatting as desired
\title{Two-Stage LASSO ADMM Signal Detection Algorithm For Large Scale MIMO}
% author names and affiliations
% use a multiple column layout for up to three different
%2-Stage LASSO ADMM For Large Scale MIMO Detection}
% affiliations
\author{\IEEEauthorblockN{}
\IEEEauthorblockA{Anis Elgabli, Ali Elghariani*, Abubakr O. Al-Abbasi, and Mark Bell\\
School of Electrical and Computer Engineering
Purdue University, West Lafayette IN 47907\\
* University of Tripoli, Libya
}}
%Email: aelghari@purdue.edu and mikedz@purdue.edu }}
\maketitle
\vspace{-.4in}
\begin{abstract}
This paper explores the benefit of using some of the machine learning techniques and Big data optimization tools in approximating maximum likelihood (ML) detection of Large Scale MIMO systems. First, large scale MIMO detection problem is formulated as a LASSO (Least Absolute Shrinkage and Selection Operator) optimization problem. Then, Alternating Direction Method of Multipliers (ADMM) is considered in solving this problem. The choice of ADMM is motivated by its ability of solving convex optimization problems by breaking them into smaller sub-problems, each of which are then easier to handle. Further improvement is obtained using two stages of LASSO with interference cancellation from the first stage. The proposed algorithm is investigated at various modulation techniques with different number of antennas. It is also compared with widely used algorithms in this field. Simulation results demonstrate the efficacy of the proposed algorithm for both uncoded and coded cases.   
%In this paper, we present a low complexity MIMO detection algorithm that is based on LASSO (least absolute shrinkage and selection operator) estimation. It provides better trade-offs between complexity and performance, especially in large-scale MIMO systems. It also achieves better bit error rate (BER) performance than known heuristic algorithms in large-scale MIMO literature, such as Local Ascent Search and Reactive Tabu Search algorithms, especially at higher-order modulations. We further proposed two stage LASSO with the concept of interference cancellation and also with the concept of shadow area constraints as a measure of symbols reliability to improve the detection accuracy. We evaluate both LASSO and 2-LASSO, and we compare them with state-of-art detection methods. Numerical results demonstrate the efficacy of the proposed algorithm in large-scale MIMO systems for both uncoded and turbo coded cases.   
\end{abstract}

\section{Introduction}
\label{sec:intro}
Large scale Multi Input Multi Output antennas (MIMO) is a key technology for various wireless systems, because of its numerous advantages in providing high capacity and signal quality \cite{boccardi2013five}. It has applications for both single user and multi user wireless systems, however, a single user MIMO with a large number of antennas at both sides of the wireless link gained more interest nowadays, especially in the next generation 5G wireless backhaul links \cite{zhang2015large}.\\
\indent The challenges in presenting large scale MIMO signal detection algorithms depend on how to achieve good quality in terms of bit error performance simultaneously with low complexity at various modulation orders. Several techniques have been proposed in the literature to address these challenges. One group of algorithms that is based on the neighborhood search technique, such as the family of Likelihood Ascent Search (LAS) \cite{vishnu2008low}, \cite{li2010multiple} and Reactive Tabu Search (RTS) \cite{rajan2009low}, \cite{srinidhi2009near}, \cite{datta2010random}. These algorithms have advantages mainly at low modulation orders, such as QPSK, however their performance and complexity deteriorates at higher modulations and higher number of antennas \cite{elghariani2016low}. Another group of algorithms that are based on semidefinite programming (SDP), such as Quadratic Programming and Branch and Bound \cite{elghariani2016low}. In addition to other algorithms that use believe propagation ideas \cite{takahashi2016normalized}. \\
\indent In this paper, we formulate ML problem into least absolute shrinkage and selection operator (LASSO) optimization problem \cite{tibshirani1996regression}. The success of LASSO  in performing high dimensional data clustering and classification \cite{elhamifar2013sparse} motivated us to cast the problem of the large-scale MIMO detection into the domain of data classification. Among many useful features, one
key feature of the this formulation is that its modular structure which allows one to sparsely represent any received signal as a linear combination of the modulated symbols.

\indent LASSO is a convex optimization problem which can be solved using any generic convex solver. However, we choose to use alternating direction method of multipliers (ADMM) because it takes the advantage of the separability of the LASSO objective function and makes the variable updates in every iteration much easier. That's, at every iteration, each variable is updated in a closed form expression by solving an unconstrained strong convex optimization problem. ADMM is a widely used algorithm for solving separable convex optimization problems with linear constraints. Its global convergence was established in the early 1990’s by Eckstein and Bertsekas \cite{Eckstein:1992:DSM:153390.153393}. The interest in ADMM has exploded in recent years because of applications in signal and image processing, compressed sensing \cite{2009arXiv0912.1185Y}, distributed optimization, statistical machine learning \cite{Boyd:2011:DOS:2185815.2185816}, and quadratic and linear programming \cite{doi:10.1137/120878951}. \\
\indent We further improve LASSO ADMM detection by implementing two stages LASSO optimization for better performance, which will be denoted in this paper as (2 LASSO ADMM). The idea is to implement the second stage of LASSO ADMM detection based on interference cancellation from the first stage. This improves the detection performance significantly. The simulation experiments show the efficacy of the proposed algorithm in both coded and uncoded performance at various QAM modulations.\\
\indent The remainder of this paper is organized as follows. In Section II, system model is discribed. Section III contains problem formulation, and section IV presents proposed algorithm. Finally, simulation results and conlusion are provided.

\section{System Model}
We consider a MIMO system with $N_t$ transmit antennas and $ N_r $ receive antennas employing a spatial multiplexing (V-BLAST) transmission. At the transmitter side,  the source information is generated  and then mapped to symbols of different alphabet. The mapped complex symbols are demultiplexed into $N_t$ separate independent data streams with a transmitted signal vector 
$\tilde{\textbf{x}}=[{\tilde{x}}_{1},\dots,{\tilde{x}}_{N_t}]^{T} \in \mathbb{C}^{N_t \times 1}$. The general MIMO channel model is :
\begin{equation}
\tilde{\textbf{{y}}}=\tilde{\textbf{{H}}}\tilde{\textbf{{x}}}+\tilde{\textbf{{n}}}
\label{eq: 1}
\end{equation}
where ${\tilde{\textbf{y}}}=[{\tilde{y}}_{1}, \dots, {\tilde{y}}_{N_r}]^{T} \in \mathbb{C}^{N_r\times1}$ is the received signal vector at all $N_r$ antennas, $\tilde{\textbf{H}} \in \mathbb{C}^{N_r\times N_t}$ denotes the flat fading  channel gain matrix whose entries are modeled as $ \mathbb{C}\mathcal{N}(0, 1) $, and $\tilde{\textbf{{n}}}$ represents the receiver AWGN noise vector whose entries are modeled as i.i.d $ \mathbb{C}\mathcal{N}(0, \sigma^2)$. The tilde symbol in (\ref{eq: 1}) is made to distinguish the complex model from the real model. We assume ideal channel estimation and synchronization at the receiver end. The ML problem of model (\ref{eq: 1}), which is equivalent to Euclidean distance minimization, can be expressed as: 
\small
\begin{equation}
\widehat{\widetilde{\textbf{{x}}}}= \underset{{{{\widetilde{{\textbf{x}}}}}\in\widetilde{\chi}^{N_t}}}{\text{argmin}}      \parallel{\widetilde{\textbf{y}}-\widetilde{\textbf{{H}}\,}\widetilde{\textbf{{x}}}\parallel_2^{2}}
\label{eq: 2}
\end{equation}
\normalsize
where $ \tilde{\chi}^{N_t} \normalsize$ is the set of all possible $N_t$-dimensional complex candidate vectors of the transmitted vector $ \tilde{\textbf{{x}}}$. The equivalent real system model of (\ref{eq: 1}) is:
\small
\begin{equation}
	\vspace{-.05in}
\textbf{{y}}=\textbf{{H}}\textbf{{x}}+\textbf{{v}}
\label{eq: 3}
	\vspace{-.00in}
\end{equation}
\begin{equation}
{\textbf{{y}}}=\begin{array}{c}
\begin{bmatrix}\Re\{{\tilde{\textbf{{y}}}}\}\\ \Im\{{\tilde{\textbf{{y}}}}\}\end{bmatrix}\end{array}, \textbf{{x}}=\begin{array}{c} \begin{bmatrix}\Re\{{{\tilde{\textbf{{x}}}}}\}\\ \Im\{{{\tilde{\textbf{{x}}}}}\}\end{bmatrix}\end{array}, \textbf{{n}}=\begin{array}{c}
\begin{bmatrix}\Re\{{{\tilde{\textbf{{n}}}}}\}\\ \Im\{{{\tilde{\textbf{{n}}}}}\}\end{bmatrix}\end{array}\\
\label{eq: 4}
\end{equation}
\begin{equation}
\mathbf{{H}}=\begin{bmatrix}\Re\{\tilde{\textbf{{H}}}\} & -\Im\{\tilde{\textbf{{H}}}\}\\ \Im\{\tilde{\textbf{{H}}}\} & \Re\{\tilde{\textbf{{H}}}\}\end{bmatrix} 
\label{eq: 5}
\end{equation}
\normalsize
In this real-valued system model, the real part of the complex data symbols is mapped to
$ [x_1,\dots, x_{Nt} ] $ and the imaginary part of these symbols is mapped to $ [x_{Nt+1},\dots, x_{2Nt} ] $. Now, the equivalent ML detection problem of the real model is :
\small
\begin{equation}
\widehat{{\textbf{x}}}=  \underset{{{\textbf{x}}\in\chi^{2N_t}}}{\text{argmin}} \parallel{{\textbf{y}}}-{{\textbf{H}}}{{\textbf{x}}}\parallel_2^{2}
\label{eq: 6}
\end{equation}
\normalsize
where set $ \chi=\{-\sqrt{\textit{M}}+1,..,-1,1,...,\sqrt{\textit{M}}-1\}$, and $\textit{M} \normalsize$ is the QAM constellation size.\\
%Each element of this real set can be transformed to a positive integer using the following linear transformation: $\textbf{{z}}=\frac{\textbf{{x}}+(\sqrt{\textit{C}}-1)}{2}$. Norm 2 term in (\ref{eq: 6}) can be simplified and the ML problem can be reformed as:

\section{Problem Formulation}

To introduce our formulation, we first start by representing each symbol in the transmitted vector $\textbf{x}$ as a sparse linear combination of the elements in $\chi$. To illustrate, if the constellation size is $M$, there will be $m=\sqrt M$ elements, {$s_1, s_2, ...., s_{m}$}. Hence, $\textbf{x}$ can be expressed as
 	\vspace{-.10in}
\begin{align}
	\vspace{-3.5in}
\textbf{x} & =\boldsymbol{S}\boldsymbol{\alpha}
 	\vspace{-2.25in}
\end{align}

where

$\!\!\ensuremath{\boldsymbol{S\!=}\!\left(\!\!\begin{array}{cccccccccc}
s_{1} & \ldots & s_{m} & 0 & \ldots &  &  &  &  & 0\\
0 & \ldots & 0 & s_{1} & \ldots & s_{m} & 0 &  &  & 0\\
\vdots\! &  & \! &  & \! &  & \ddots\\
 &  &  &  &  &  &  & \ddots\\
0 & \ldots & \ldots & \ldots &  &  &  & s_{1} & \ldots & s_{m}
\end{array}\!\!\right)}$

and,
\\
$\boldsymbol{\alpha=}\left(\begin{array}{cccccccc}
\alpha_{1} & \ldots & \alpha_{m} & \alpha_{m+1} & \ldots & \alpha_{2m+1} & \ldots & \alpha_{N_{t}m}\end{array}\right)^{T}$

Note that the first transmitted symbol is expressed as follows: 
\begin{equation}
\textbf{x}(1)=s_{1}\alpha_{1}+s_{2}\alpha_{2}+\ldots+s_{n}\alpha_{n}+\ldots+s_{m}\alpha_{m}. 
\end{equation}
Similarly, the second symbol can be written as
\begin{equation}
\textbf{x}(2)=s_{1}\alpha_{(m+1)}+s_{2}\alpha_{(m+2)}+\ldots+s_{m}\alpha_{2m}.
\end{equation}
In general, the $i$th symbol can be  expressed as
\begin{equation}
\textbf{x}(i)=s_{1}\alpha_{((i-1)m+1)}+s_{2}\alpha_{((i-1)m+2)}+\ldots+s_{m}\alpha_{im}
\end{equation}
We denote the coefficients of the element \textbf{x}(i) $\{\alpha_{((i-1)m)},.....\alpha_{im}\} \in \boldsymbol{\alpha}$ by ${\boldsymbol{\alpha}_x}_{i}$. 
%We denote the coefficients of the element \textbf{x}(i), i.e., ${\boldsymbol{\alpha}_x}_{i} \triangleq$
%$\{\alpha_{((i-1)m)},.....\alpha_{im}\} \in \boldsymbol{\alpha}$ by . 
Typically, if $\textbf{x}(i)=s_n$, then the only non zero element $\in {\boldsymbol{\alpha}_x}_{i}$ is $\alpha_{(i-1)m+n}$, and its value is exactly one. Therefore, the following holds true
 % i is the only non-zero entry whose value is exactly equal to one in the range $\{\alpha_{(i-1)m},.....\alpha_{im}\}$ i.e ( $\alpha_{((i-1))m+n}=1$, and ${\boldsymbol{\alpha}_x}_{i}\\\alpha_{((i-1)m+n)}=0$). Therefore, we have
 	\vspace{-.10in}
\begin{equation}
 	\vspace{-.10in}
\sum_{j=(i-1)m}^{im}\alpha_{j}=1, \forall i \label{sum_alpha}
 	\vspace{-.00in}
\end{equation}

We can write \eqref{sum_alpha} in a matrix form as follow 
 	\vspace{-.1in}
\begin{equation}
\boldsymbol{B}\boldsymbol{\alpha}=\boldsymbol{1}
\end{equation}
 	\vspace{-0.2in}
where

\[
\boldsymbol{B=}\left(\begin{array}{cccccccccc}
1 & \ldots & 1 & 0 & \ldots &  &  &  &  & 0\\
0 & \ldots & 0 & 1 & \ldots & 1 & 0 &  &  & 0\\
\vdots &  &  &  &  &  & \ddots\\
 &  &  &  &  &  &  & \ddots\\
0 & \ldots & \ldots & \ldots &  &  &  & 1 & \ldots & 1
\end{array}\right)
\]

 Next, we explain our MIMO detection optimization framework. Mathematically, we formulate our constrained optimization problem as follows:

%\begin{equation}
%\begin{array}{cc}
%\widehat{\boldsymbol{\text{x}}}=\underset{\text{\ensuremath{\boldsymbol{x}\in}}\mathbb{R}^{mN_{t}}}{\text{argmin}} & \left(\left\Vert \boldsymbol{\boldsymbol{\text{y}-\text{Hx}}}\right\Vert _{2}^{2}+\lambda\left\Vert \boldsymbol{\alpha}\right\Vert _{0}\right)\\
%\text{subject to}\\
% & \boldsymbol{\text{\ensuremath{\boldsymbol{x}}-}\boldsymbol{S}\alpha=1}\\
% & \boldsymbol{\boldsymbol{B}}\boldsymbol{\alpha}=1
%\end{array}\label{eq:first_form}
%\end{equation}

	\vspace{-0.2in}
\begin{equation}
\begin{array}{cc}
\widehat{\textbf{x}}=\underset{\text{\ensuremath{\textbf{x}\in}}\mathbb{R}^{mN_{t}}}{\text{argmin}} & \left(\left\Vert \boldsymbol{\boldsymbol{\textbf{y}-\text{\textbf{H}\textbf{x}}}}\right\Vert _{2}^{2}+\lambda\left\Vert \boldsymbol{\alpha}\right\Vert _{0}\right)\\
\text{subject to}\\
 & \boldsymbol{\text{\ensuremath{\textbf{x}}-}\boldsymbol{S}\alpha=0}\\
 & \boldsymbol{\boldsymbol{B}}\boldsymbol{\alpha}=1
\end{array}\label{eq:first_form}
\end{equation}
	\vspace{-0.05in}
In \eqref{eq:first_form}, the constraint of the non convexity of the feasible set on $\textbf{x}$ is relaxed, but norm zero in the objective function and the two constraints are introduced to approximate the solution of the problem given in (6) if $\lambda$ is chosen carefully.   

%Here, we approximate the problem of minimizing the quadratic term $\parallel{{\textbf{y}}}-{{\textbf{H}}}{{\textbf{x}}}\parallel_2^{2}$ subject to the discrete feasible set of the transmitted symbols by solving \eqref{eq:first_form}. Further, we relax the discretization constraint on the elements of $\boldsymbol{\alpha}$, and $\textbf{x}$. Note that $\boldsymbol{\alpha}\in \{0,1\}$, and every element in $\textbf{x} \in \{s_1,....,s_n\}$. 

However, the formulation in \eqref{eq:first_form} has two issues. First, minimizing the zero norm is a well known NP hard problem. Second, compared to problem (6) another variable ($\alpha$) is introduced which increases the number of optimized variables. 

%rather than optimizing over $\boldsymbol{x}$ only, both $\textbf{x}$ and $\boldsymbol{\alpha}%$ need to be optimized. 

In order to solve the first issue, we replace norm zero by norm one (convex relaxation of norm 0), hence the problem becomes convex. Moreover, the second issue can be easily resolved by expressing $\textbf{x}$ in terms of $\boldsymbol{\alpha}$ and that will further cancel the first constraint.  
%hence the optimization is turned out to be over a single variable $\boldsymbol{\alpha}$. Thus,  
so, \eqref{eq:first_form} reduces to:

	\vspace{-0.1in}
%\small
\begin{equation}
\hat{{\boldsymbol{\alpha}}}=  \underset{{{\boldsymbol{\alpha}}}}{\text{argmin}} \bigg(\frac{1}{2}\parallel{{\textbf{y}}}-{\widetilde{\textbf{H}}}{{\boldsymbol{\alpha}}}\parallel_2^{2}+\lambda^\prime ||\boldsymbol{\alpha}||_1 \bigg)
\label{conv_relax}
\end{equation}
%\normalsize
\,\,\,\,\,\,\, subject to\\
\begin{equation}
\textbf{B}\boldsymbol{\alpha}=\textbf{1}.\nonumber
%\label{conv_relax_con}
\end{equation}
%We denote $\widetilde{\boldsymbol{H}}=\textbf{HS}$, and $\lambda^\prime=\frac{\lambda}{2}$, 
where $\widetilde{\textbf{H}}=\textbf{H\,S}$, and $\lambda^\prime=\frac{\lambda}{2}$. Note that \eqref{conv_relax} is constrained LASSO. However, since the constraint in (\ref{conv_relax}) can tolerate some violation, it can be expressed as non-constrained optimization problem as follows:
	\vspace{-0.0in}
\begin{equation}
\hat{{\boldsymbol{\alpha}}}=  \underset{{{\boldsymbol{\alpha}}}}{\text{argmin}} \bigg(\frac{1}{2}\parallel{{\textbf{y}}}-{{\textbf{$\widetilde{\boldsymbol{H}}$}}}{{\boldsymbol{\alpha}}}\parallel_2^{2}+\mu \parallel{{\textbf{1}}}-{{\textbf{B}}}{{\boldsymbol{\alpha}}}\parallel_2^{2}+\lambda^\prime ||\boldsymbol{\alpha}||_1 \bigg).
\label{eq:unconRelax}
\end{equation}
\normalsize
We choose  $\mu$ such that $\mu \gg \text{max}(\frac{1}{2},\lambda^\prime)$ to prioritize reducing the constraint violation over fitting the quadratic term and  norm one minimization. Finally, \eqref{eq:unconRelax} can be re-written in the standard LASSO form as follow as follows:
\begin{equation}
\widehat{\boldsymbol{\alpha}}=\underset{\boldsymbol{\alpha}}{\text{argmin}}\bigg(\frac{1}{2}\parallel\overline{\textbf{y}}-\overline{\textbf{H}}\,\boldsymbol{\alpha}\parallel_{2}^{2}+\lambda^{\prime}||\boldsymbol{\alpha}||_{1}\bigg)
\label{final_form_lasso}
\end{equation}
\normalsize
% , so equation (\ref{eq: moderate}) will reduce to:
%\begin{equation}
%\hat{{\textbf{$\alpha$}}}=  \underset{{{\textbf{$\alpha$}}}}{\text{argmin}} \bigg(\parallel{\widetilde{\boldsymbol{y}}}-\hat{{\textbf{H}}}{{\textbf{$\alpha$}}}\parallel_2^{2}+\lambda |\textbf{$\alpha$}|_1 \bigg)
%\label{eq: lasso1}
%\end{equation}
%\normalsize
where
%$\widetilde{\boldsymbol{y}}=\left[\begin{array}{cc}
%\boldsymbol{y} & \boldsymbol{1}\end{array}\right]$ and $\widetilde{\boldsymbol{H}}=\mu^{\frac{1}{2}}\left[\begin{array}{cc}
%\boldsymbol{H} & \boldsymbol{B}\end{array}\right]$.

 $\overline{\textbf{y}}=\begin{bmatrix}
\textbf{y}\\
\sqrt{\mu}\,\textbf{1}
\end{bmatrix}$, and $\overline{\textbf{H}}=
\begin{bmatrix}
\textbf{$\widetilde{\boldsymbol{H}}$}\\
\sqrt{\mu}\,\textbf{B}
\end{bmatrix}
 $
 
% Finally, (\ref{eq: lasso1}) can be expressed compactly as below:
% \begin{equation}
%\hat{{\boldsymbol{\alpha}}}=  \underset{{{\boldsymbol{\alpha}}}}{\text{argmin}} \bigg(\frac{1}{2}\parallel{{\textbf{y$^\prime$}}}-\hat{{\textbf{H}}}{{\boldsymbol{\alpha}}}\parallel_2^{2}+\lambda^\prime |\boldsymbol{\alpha}|_1 \bigg)
%\label{eq: lasso}
%\end{equation}
%\normalsize

%This transformation from constrained to unconstrained LASSO turns out to be quite accurate as will be shown in the numerical results. 
%The optimization problem given in \eqref{final_form_lasso} can be solved using any convex optimization solver. Practically, the adopted optimization technique should be fast, and guaranteed convergence and those motivated our choice of the ADMM. 
Next, we present our ADMM based algorithm to  solve this problem efficiently.
\section{Proposed Algorithm}

\subsection{LASSO-ADMM}
%LASSO is well known problem in machine learning and big data optimization. It is used in problems of high dimensional data clustering and classification. For example, if we are given a point that falls into a union of subspaces, and we need to classify to which subspace it belongs,  we represent it as a linear combination of other points and solve LASSO problem to find the most sparse combination which will be from the points that fall in the same subspace \ref{}.

 To find the coefficients vector ($\boldsymbol{\alpha}$), we solve \eqref{final_form_lasso} using ADMM framework described in Algorithm 1. ADMM convergence is proven for any problem with an objective function that is sum of two separable convex functions with linear constraints. Hence, we re-formulate LASSO as a sum of two separable convex functions with a linear constraint as follows:   

\begin{equation}
\underset{{{\boldsymbol{\alpha},\textbf{z}}}}{\text{min}} \bigg(\frac{1}{2}\parallel{\overline{\textbf{y}}}-\overline{{\textbf{H}}}{{\boldsymbol{\alpha}}}\parallel_2^{2}+\lambda^\prime ||\textbf{z}||_1 \bigg)
\label{eq:lassoAdmm}
\end{equation}
subject to
\begin{equation}
\boldsymbol{\alpha}=\textbf{z}
\label{eq:lassoAdmm_con1} 
\end{equation}
\normalsize
Note that the constrained optimization problem (\ref{eq:lassoAdmm}-\ref{eq:lassoAdmm_con1}) is equivalent to \eqref{final_form_lasso}

%we replaced $\boldsymbol{\alpha}$ by $\boldsymbol{\textbf{z}}$ in the norm one , and  introduced the constraint given in  \eqref{eq:lassoAdmm_con1}.

The augmented Lagrangian for \eqref{eq:lassoAdmm}-\eqref{eq:lassoAdmm_con1} is:
\begin{align}
\mathcal{\boldsymbol{\mathcal{L}}}_{\rho}\left(\boldsymbol{\alpha},\,{\textbf z},\,\boldsymbol{\eta}\right) & =\frac{1}{2}\parallel\overline{\textbf{y}}-\overline{\textbf{H}}\boldsymbol{\alpha}\parallel_{2}^{2}+\lambda^{\prime}||{\textbf z}||_{1}+\boldsymbol{\eta}^{T}\left(\boldsymbol{\alpha}-{\textbf z}\right)\nonumber \\
 & \,\,\,\,\,\,\,\,+\frac{\rho}{2}\parallel\boldsymbol{\alpha}-{\textbf z}\parallel_{2}^{2}
\end{align}

At the $(k+1)$th iteration, the primal ($\boldsymbol{\alpha},{\bf z}$) and dual ($\boldsymbol{\eta}$) variables are updated sequentially as follows:
%\begin{equation}
%\left\{\begin{array}{l}
%\boldsymbol{\alpha}^{k+1}= \underset{{{\boldsymbol{\alpha}}}}{\text{argmin}} \mathcal{L}_\rho(\boldsymbol{\alpha},\textbf{z}^{k},\textbf{$\eta$}^{k})\\
%\boldsymbol{\text{z}}^{k+1}= \underset{{{\textbf{z}}}}{\text{argmin}} \mathcal{L}_\rho(\textbf{$\alpha$}^{k+1},\textbf{z},\textbf{$\eta$}^{k})\\
%\eta^{k+1}= \eta^k+(\alpha^{k+1}-\textbf{z}^{k+1})
%\end{array}\right.
%\end{equation}
\begin{align}
\boldsymbol{\alpha}^{k+1} & =\underset{\boldsymbol{\alpha}}{\text{argmin}}\mathcal{\boldsymbol{\mathcal{L}}}_{\rho}\left(\boldsymbol{\alpha},\,{\textbf z}^{k},\,\boldsymbol{\eta}^{k}\right)\\
\boldsymbol{\textbf z}^{k+1} & =\underset{{\textbf z}}{\text{argmin}}\mathcal{\boldsymbol{\mathcal{L}}}_{\rho}\left(\boldsymbol{\alpha}^{k+1},\,{\textbf z},\,\boldsymbol{\eta}^{k}\right)\\
\boldsymbol{\eta}^{k+1} & =\boldsymbol{\eta}^{k}+\left(\boldsymbol{\alpha}^{k+1}-{\textbf z}^{k+1}\right)
\end{align}

The primal variables ($\boldsymbol{\alpha}$, $\textbf{z}$) are updated as follows:
%In order to describe the Algorithm, we first define the two sub-problems:
\begin{itemize}
\item {\bf $\boldsymbol{\alpha}$-update}:
%\begin{equation}
%\label{eq: alphaUpdate}
%\end{equation}
\vspace{-0.10in}
\begin{equation}
\boldsymbol{\alpha}^{k+1}=\underset{\boldsymbol{\alpha}}{\text{argmin}}\bigg(\frac{1}{2}\parallel\overline{\textbf{y}}-\overline{\textbf{H}}\boldsymbol{\alpha}\parallel_{2}^{2}+\left(\boldsymbol{\eta}^{k}\right)^{T}\boldsymbol{\alpha}+\frac{\rho}{2}\parallel\boldsymbol{\alpha}-{\textbf z}\parallel_{2}^{2}\bigg)
 \label{eq: alphaUpdate}
\end{equation}

%\begin{equation}
%\alpha^{k+1}= \underset{{{\textbf{$\alpha$}}}}{\text{argmin}} \Big(\frac{1}{2}\parallel{{\textbf{y$^\prime$}}}-\hat{{\textbf{H}}}{{\textbf{$\alpha$}}}\parallel_2^{2}+\textbf{$\eta$}^T\textbf{$\alpha$}+\rho(\frac{1}{2} \alpha^T\alpha-\textbf{Z}^T\alpha)\Big)
%\end{equation}
Since \eqref{eq: alphaUpdate} is strictly convex with respect to $\boldsymbol{\alpha}$, taking the derivative and equating to zero yields a closed form expression for obtaining $\boldsymbol{\alpha}^{k+1}$.

\item {\bf z-update}:
\vspace{-0.050in}
%\begin{equation}
%Z^{k+1}= \underset{{{\textbf{Z}}}}{\text{argmin}} \Big(\lambda^\prime |\textbf{Z}|_1-{\textbf{$\eta$}^T}^{k}\textbf{Z}+\frac{\rho}{2}\parallel \textbf{$\alpha$}^{k+1}-\textbf{Z} \parallel_2^2\Big)
%\label{eq: zUpdate}
%\end{equation}
\vspace{-0.10in}
\begin{equation}
{\textbf z}^{k+1}=\underset{\textbf z}{\text{argmin}}\bigg(\lambda^{\prime}||{\textbf z}||_{1}-\left(\boldsymbol{\eta}^{k}\right)^{T}{\textbf z}+\frac{\rho}{2}\parallel\boldsymbol{\alpha}^{k+1}-{\textbf z}\parallel_{2}^{2}\bigg)
\label{zUpdate}
\end{equation}

Since the minimized function in (\ref{zUpdate}) is separable with respect to every element in {\textbf z},  it can be written as a sum of independent terms: 
\vspace{-0.10in}
\begin{equation}
f(\boldsymbol{\alpha}^{k+1},\textbf{z},\boldsymbol{\eta}^{k})=\sum_{i=1}^{N}\Big(\frac{\rho}{2}(\textbf{\ensuremath{\boldsymbol{\alpha}}}_{i}^{k+1}-\textbf{z}_{i})^{2}+\lambda^{\prime}|\textbf{z}_i|-{\eta^{k}}_{i}\textbf{z}_{i}\Big)
\end{equation}
For each element, $\textbf{z}_i^{k+1}$, the solution can therefore be computed by independently
minimizing $f(\boldsymbol{\alpha}^{k+1},\textbf{z},\boldsymbol{\eta}^{k})$ with respect to that element. Moreover, all elements can be updated in parallel. Hence, the $i$th element update is defined by the following equation: 
%\begin{align*}
%&\underset{{{\textbf{Z}}}}{\text{argmin}}\left\{\begin{array}{l}
%\frac{\rho}{2}(\textbf{$\alpha$}_i^{k+1}-\textbf{Z}_i)^2+\lambda^\prime |\textbf{Z}|_i-{\eta^T}_i^k \textbf{Z}_i
%\end{array}\right\}
%\\
%&=\left\{\begin{array}{l}\frac{1}{\rho}(\rho(\alpha_i^{k+1}+\eta_i^{k})-\lambda^\prime)  \quad  \text{for} \quad \rho(\alpha_i^{k+1}+\eta_i^{k}) > \lambda^\prime\\
%0  \quad \quad \quad \quad \quad \quad \quad \quad \quad \quad \text{for} \quad |\rho(\alpha_i^{k+1}+\eta_i^{k})| \leq \lambda^\prime\\
%\frac{1}{\rho}(\rho(\alpha_i^{k+1}+\eta_i^{k})+\lambda^\prime)  \quad \text{for} \quad \rho(\alpha_i^{k+1}+\eta_i^{k}) < \lambda^\prime\\
%\end{array}\right\}
%\label{eq: zUpdate2}
%\end{align*}
\vspace{-0.050in}
\begin{align}
\underset{{\textbf{z}_i}}{\text{argmin}}\left\{ \begin{array}{l}
\frac{\rho}{2}(\textbf{\ensuremath{\alpha}}_{i}^{k+1}-\textbf{z}_{i})^{2}+\lambda^{\prime}|\textbf{z}_i|-{\eta}_{i}^{k}\textbf{z}_{i}\end{array}\right\} =\nonumber \\
\begin{cases}
\begin{array}{l}
\frac{1}{\rho}(\rho(\alpha_{i}^{k+1}+\eta_{i}^{k})-\lambda^{\prime}),\end{array} & \rho(\alpha_{i}^{k+1}+\eta_{i}^{k})>\lambda^{\prime}\\
0 & |\rho(\alpha_{i}^{k+1}+\eta_{i}^{k})|\leq\lambda^{\prime}\\
\frac{1}{\rho}(\rho(\alpha_{i}^{k+1}+\eta_{i}^{k})+\lambda^{\prime}) & \rho(\alpha_{i}^{k+1}+\eta_{i}^{k})<-\lambda^{\prime}
\end{cases}
\end{align}

\end{itemize}

%We now see the advantage of ADMM in solving LASSO. The two functions that form the objective functions are separable and can be solved by different techniques. However, ADMM is taking the advantage of the separability structure of the two functions, and individually updates $\boldsymbol{\alpha}$ from updating $\boldsymbol{z}$. 

%Updating $\boldsymbol{\alpha}$ is just solving unconstrained strictly convex function which yields closed form expression. Similarly, updating $\boldsymbol{z}$ is Shrinkage operator  which is also a closed form expression. Then, both updated variables are used to update the dual variable of the next iteration. 

 After the coefficients ($\boldsymbol{\alpha}$ vector) are found, the received vector {\bf x} can be found, $\textbf{x}=\boldsymbol{S\alpha}$. Further, since ${\boldsymbol{\alpha_x}}_i$ may not be sparse for every $\textbf{x}(i)$, $\textbf{x}(i)$ is quantized to the nearest symbol in the constellation set $ \chi$, (lines 10-12).  The quantization function is defined by $\boldsymbol{Q}(\textbf{x}(i),\chi)$.
 
 %we split the line (decision region) between any two modulated symbols into two regimes based on a predefined threshold. Then, we quantize any value in the lower regime (lies in the left to the threshold) to the lower symbol $s_l$, and any point that falls in the higher region (lies to the right of the threshold) is quantized to the higher symbol $s_h$ 
%(lines 10-12).  The quantization function is defined by $\boldsymbol{Q}(\textbf{x}(i),\chi)$.

%Once the coefficient vector $\boldsymbol{\alpha}$ is found, the coefficients the estimated symbol vector is found by substituting in equation \ref{eq: con1}. i.e $\hat{\textbf{x}}=\textbf{S}^{-1}\alpha$
\begin{figure}

		%\vspace{-.6in}
		\begin{minipage}{\linewidth}
			\begin{algorithm}[H]
					{\tiny 
				\small
				\begin{algorithmic}[1]
				    \STATE {\bf 1: Initilization}
    
   \STATE $(\boldsymbol{\bf \alpha}^0, \textbf{z}^0, \boldsymbol{\bf \eta}^0)\leftarrow(\textbf{0,0,0})$
   \STATE $k=1$
   %\STATE  $\boldsymbol{\bf \eta}^1 \leftarrow \infty$
    \STATE {\bf 2: ADMM iteration:}
    \WHILE{(1)}
  % \WHILE{$|\ensuremath{|\boldsymbol{\eta}^{k}-\boldsymbol{\eta}^{k-1}||>\epsilon}$}
   \STATE $\ensuremath{\boldsymbol{\alpha}^{k}\leftarrow\underset{\ensuremath{\boldsymbol{\alpha}}}{\text{argmin}}\mathcal{L}_{\rho}(\textbf{\ensuremath{\boldsymbol{\alpha}}},\textbf{z}^{k-1},\textbf{\ensuremath{\boldsymbol{\eta}}}^{k-1})}$
    %\STATE $\alpha^{k+1}\leftarrow (\ref{eq: alphaUpdate})$
\STATE $\ensuremath{\boldsymbol{z}^{k}\leftarrow\underset{\ensuremath{\boldsymbol{z}}}{\text{argmin}}\mathcal{L}_{\rho}(\textbf{\ensuremath{\boldsymbol{\alpha}^{k}}},\textbf{z},\textbf{\ensuremath{\boldsymbol{\eta}}}^{k-1})}$
\STATE $\ensuremath{\boldsymbol{\eta}^{k}\leftarrow\boldsymbol{\eta}^{k-1}+(\boldsymbol{\alpha}^{k}-\textbf{z}^{k})}$
%\STATE {\bf UNTIL} $|\ensuremath{|\boldsymbol{\eta}^{k+1}-\boldsymbol{\eta}^{k}||<\epsilon}$
\IF{$|\ensuremath{|\boldsymbol{\eta}^{k}-\boldsymbol{\eta}^{k-1}||<\epsilon}$}
\STATE break
\ENDIF
\STATE $k=k+1$
\ENDWHILE

\STATE $\textbf{x}=\boldsymbol{S\,\alpha}$
\FOR{$i=1:2N_t$}
\STATE $\hat{\textbf{x}}(i)=\boldsymbol{Q}(\textbf{x}(i),\chi)$ 
%\IF{$(abs(\alpha(i)-1) \leq \delta)$}
	%\STATE $\alpha(i)=1$
	%\ELSIF{$(abs(\alpha(i)-0) \leq \delta)$}
	%\STATE $\alpha(i)=0$
	%\STATE $B(i)=0$
	%\ENDIF
\ENDFOR
   				\end{algorithmic}
				\caption{LASSO-ADMM Algorithm \label{alg1}}
}						
			\end{algorithm}
		\end{minipage}
		\vspace{-.2in}
	\end{figure}

\subsection{Two-LASSO-ADMM}
The idea of this algorithm (Algorithm \ref{alg2}) is to implement two stages of the LASSO detection with interference cancellation to further improve the detection of the unreliable symbols. The main difference compared to LASSO, is that after the first shot of LASSO, the estimate of the received vector $\textbf{x}$ is found, and the line (decision region) between any two modulated symbols ($s_l$, and $s_h$) is split into three regimes. Any element in the received vector $\textbf{x}(i)$ falls within a predefined threshold ($\tau$) from either $s_l$ or $s_r$ is rounded to that point. However, any point falls between these two points but it is not within $\tau$ distance from any of them (gray zone) 
 %determined by two thresholds: lower ($\tau_l$) and higher ($\tau_h$). 
 %Any value lower than $\tau_l$ is quantized to the lower symbol $s_l$, and any value higher than $\tau_h$ is quantized to the higher symbol $s_h$. However, any point falls in the region between $\tau_l$ and $\tau_h$ 
 is not rounded to any of the symbols, and it is referred to the second stage of the detection algorithm(lines 16-27 Algorithm \ref{alg2}). Therefore, some of the received symbols are detected from the first round, but some others are deferred to the second round (not yet detected). For any symbol $\textbf{x}(i)$ that is detected from the first round, its corresponding coefficients are found (${\boldsymbol{\alpha}_x}_i$), so they are excluded (removed) from the second round along with their contribution in $\overline{\textbf{y}}$ and corresponding columns in the matrices $\textbf{B}$ and $\textbf{S}$ (those columns denoted by ${{I_\alpha}_x}_i$. At the second LASSO round, every detected symbol ($\textbf{x}(i)$) is quantized to the nearest symbol in the constellation set $ \chi$.

We conclude this section by pointing out that the main ingredient in the computations of the LASSO detector is the $\boldsymbol{\alpha}$-update where a matrix inversion of order $\mathcal{O}((m\cdot N_t)^3)$ is needed. While the Two-LASSO requires more computations, only a few symbols are refereed to the second shot, especially for medium to high SNR. % and when $ \delta $ parameter is optimized. 
This makes the computational complexity of Algorithms 1 and 2 is nearly the same.

\begin{figure}
		%\vspace{-.6in}
		\begin{minipage}{\linewidth}
			\begin{algorithm}[H]
				\small
				\begin{algorithmic}[1]
						{\small
				    \STATE {\bf Initilization:}
				    \STATE $n=1$, $(\boldsymbol{\bf \alpha}^0, \textbf{z}^0, \boldsymbol{\bf \eta}^0)\leftarrow(\textbf{0,0,0})$, $k=1$
    \FOR{$n=1:2$}
    \STATE {\bf 2: ADMM iteration:}
    \WHILE{(1)}
  % \WHILE{$|\ensuremath{|\boldsymbol{\eta}^{k}-\boldsymbol{\eta}^{k-1}||>\epsilon}$}
   \STATE $\ensuremath{\boldsymbol{\alpha}^{k}\leftarrow\underset{\ensuremath{\boldsymbol{\alpha}}}{\text{argmin}}\mathcal{L}_{\rho}(\textbf{\ensuremath{\boldsymbol{\alpha}}},\textbf{z}^{k-1},\textbf{\ensuremath{\boldsymbol{\eta}}}^{k-1})}$
    %\STATE $\alpha^{k+1}\leftarrow (\ref{eq: alphaUpdate})$
\STATE $\ensuremath{\boldsymbol{z}^{k}\leftarrow\underset{\ensuremath{\boldsymbol{z}}}{\text{argmin}}\mathcal{L}_{\rho}(\textbf{\ensuremath{\boldsymbol{\alpha}^{k}}},\textbf{z},\textbf{\ensuremath{\boldsymbol{\eta}}}^{k-1})}$
\STATE $\ensuremath{\boldsymbol{\eta}^{k}\leftarrow\boldsymbol{\eta}^{k-1}+(\boldsymbol{\alpha}^{k}-\textbf{z}^{k})}$
%\STATE {\bf UNTIL} $|\ensuremath{|\boldsymbol{\eta}^{k+1}-\boldsymbol{\eta}^{k}||<\epsilon}$
\IF{$|\ensuremath{|\boldsymbol{\eta}^{k}-\boldsymbol{\eta}^{k-1}||<\epsilon}$}
\STATE break
\ENDIF
\STATE $k=k+1$
\ENDWHILE
\IF{$n=1$}
\STATE $\textbf{x}_n=\boldsymbol{S\,\alpha}$
\FOR{$i=1:2N_t$}
%\STATE find $s_l$ and $s_r$ of {\bf x}(i)
%\STATE $[\nu_l,\nu_r]=find(\textbf{x}(i))$
\IF{$(|{\textbf{x}_n}(i)-s_l| \leq \tau)$}
	\STATE ${\textbf{x}_n}(i)=s_l$, $\ensuremath{\boldsymbol{\alpha}=\boldsymbol{\alpha}\backslash\{\boldsymbol{\alpha}_{x_{n,i}}}\}$, $\ensuremath{\boldsymbol{B}=\boldsymbol{B}\backslash\{I_{\boldsymbol{\alpha}_{x_{n,i}}}}\}$, 
	\STATE $\ensuremath{\boldsymbol{S}=\boldsymbol{S}\backslash\{I_{\boldsymbol{\alpha}_{x_{n,i}}}}\}$
	\ELSIF{$(|{\textbf{x}_n}(i)-s_r| \geq \tau)$}
	\STATE ${\textbf{x}_n}(i)=s_r$, $\ensuremath{\boldsymbol{\alpha}=\boldsymbol{\alpha}\backslash\{\boldsymbol{\alpha}_{x_{n,i}}}\}$, $\ensuremath{\boldsymbol{B}=\boldsymbol{B}\backslash\{I_{\boldsymbol{\alpha}_{x_{n,i}}}}\}$, 
	\STATE $\ensuremath{\boldsymbol{S}=\boldsymbol{S}\backslash\{I_{\boldsymbol{\alpha}_{x_{n,i}}}}\}$
	\ELSE
	\STATE ${\textbf{x}_n}(i)=0$
	
\ENDIF	
%\STATE $\hat{\textbf{x}}(i)=Q(\textbf{x}(i),\boldsymbol{S})$
%\IF{$(abs(\alpha(i)-1) \leq \delta)$}
%	\STATE $\alpha(i)=1$
%	\ELSIF{$(abs(\alpha(i)-0) \leq \delta)$}
%	\STATE $\alpha(i)=0$
%	\STATE $B(i)=0$
%	\ENDIF
	\ENDFOR

\ELSE

%\STATE $\textbf{x}=\boldsymbol{S\,\alpha}$
\STATE $j=1$
\FOR{$i=1:2N_t$}
\IF{${\textbf{x}_1}(i)=0$}
\STATE $\textbf{x}_2(j) = \boldsymbol{Q}(\textbf{x}_2(j),\chi)$
 \STATE $\textbf{x}_1(i)=\textbf{x}_2(j)$
 \STATE $j=j+1$
 \ENDIF
\ENDFOR
\ENDIF 
\ENDFOR
\STATE $\hat{\textbf{x}}(i)=\textbf{x}_1(i)$
}
	   				\end{algorithmic}
				\caption{Two-LASSO-ADMM Algorithm \label{alg2}}
			\end{algorithm}
		\end{minipage}
		\vspace{-.2in}
	\end{figure}
\vspace{-0.050in}
\section{Numerical Results}
Simulation results for an uncoded large-scale MIMO system in a block flat fading channel with $ N_t = N_r $ is shown. We assume perfect knowledge of the channel state information at the receiver. QPSK and 16QAM modulations are considered for demonstration. ADMM parameters were as follows: $\lambda=10$, $\mu=10^6$, $\rho=10$, and $\tau=0.6$. %, and the algorithm complexity will not increase with respect to the modulation order.
%We assume that every symbol can be represented as a sparse linear combination of a scaled version of the modulated symbol where we scale every symbol by $\sqrt{2}$ ($sf=\sqrt{2}$) which seems to give slightly better performance than a linear combination of the un-scaled symbol. For 16QAM, we choose $\lambda=sf=5$, and $\tau=0.25$. Moreover, we test for the case when we assumed that every element in $\textbf{x}$ can be represented as a linear combination from the two extreme points (e,g $\{-3, 3\}$ for 16QAM) and we noticed almost the same results which dictates that the same performance could be achieved with lower complexity ($\mathcal{O}((N_t)^3)$ only.

\begin{figure*}[htb]
	%	\begin{figure*}[htb]
%		\centering
	\begin{minipage}{.50\textwidth}
%				\centering
		\includegraphics[trim=0.0in 0.00in 0.0in 0.0in, clip, width=\textwidth]{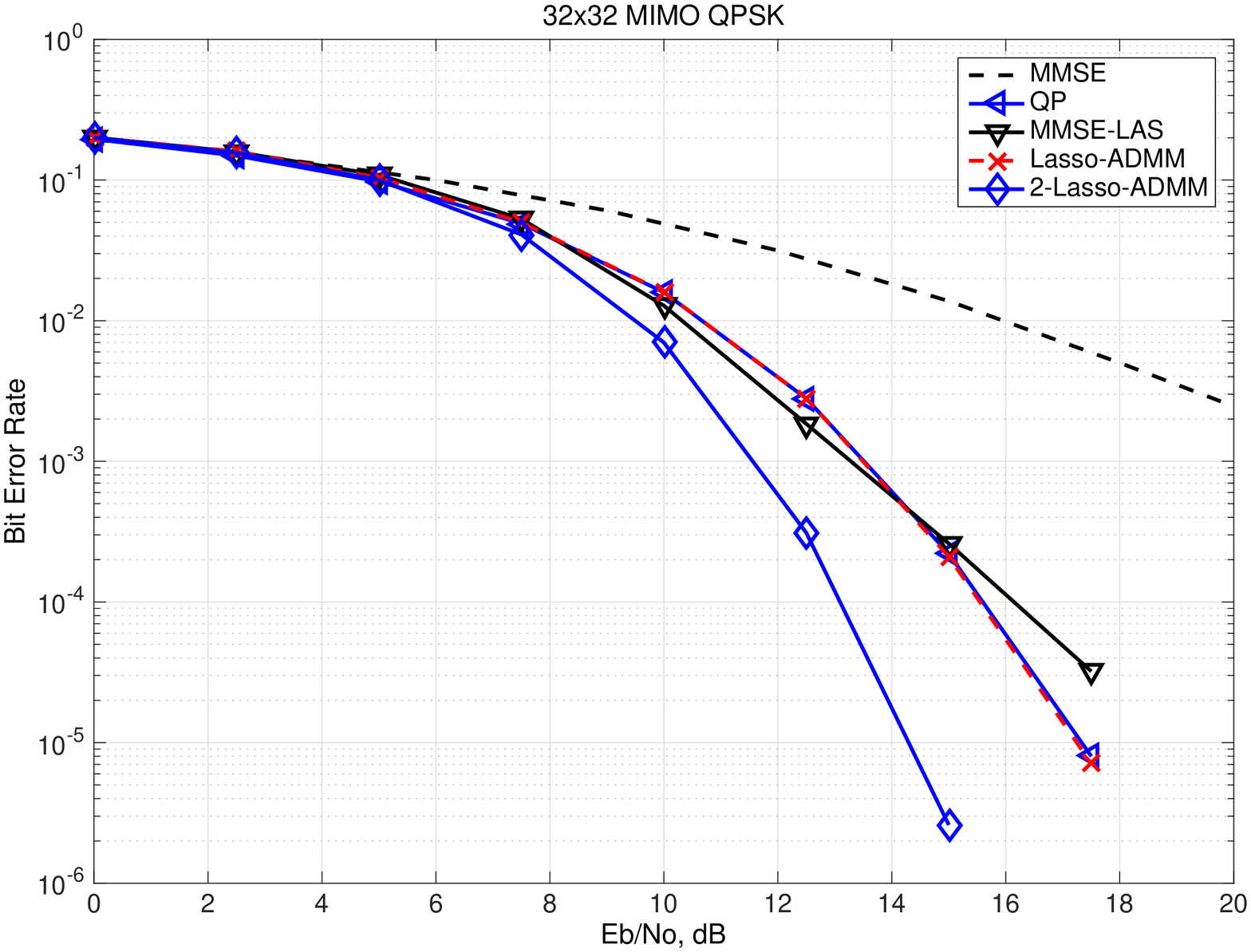}
%		\vspace{-.25in}
		\caption{\small BER performance of QPSK $32\times32$ MIMO }
		\label{fig: 1}
	\end{minipage}%
%	\hspace{0.5mm}
	\begin{minipage}{.40\textwidth}
%				\centering
		\includegraphics[trim=0.0in 0.0in 0.20in 0.0in, clip, width=\textwidth]{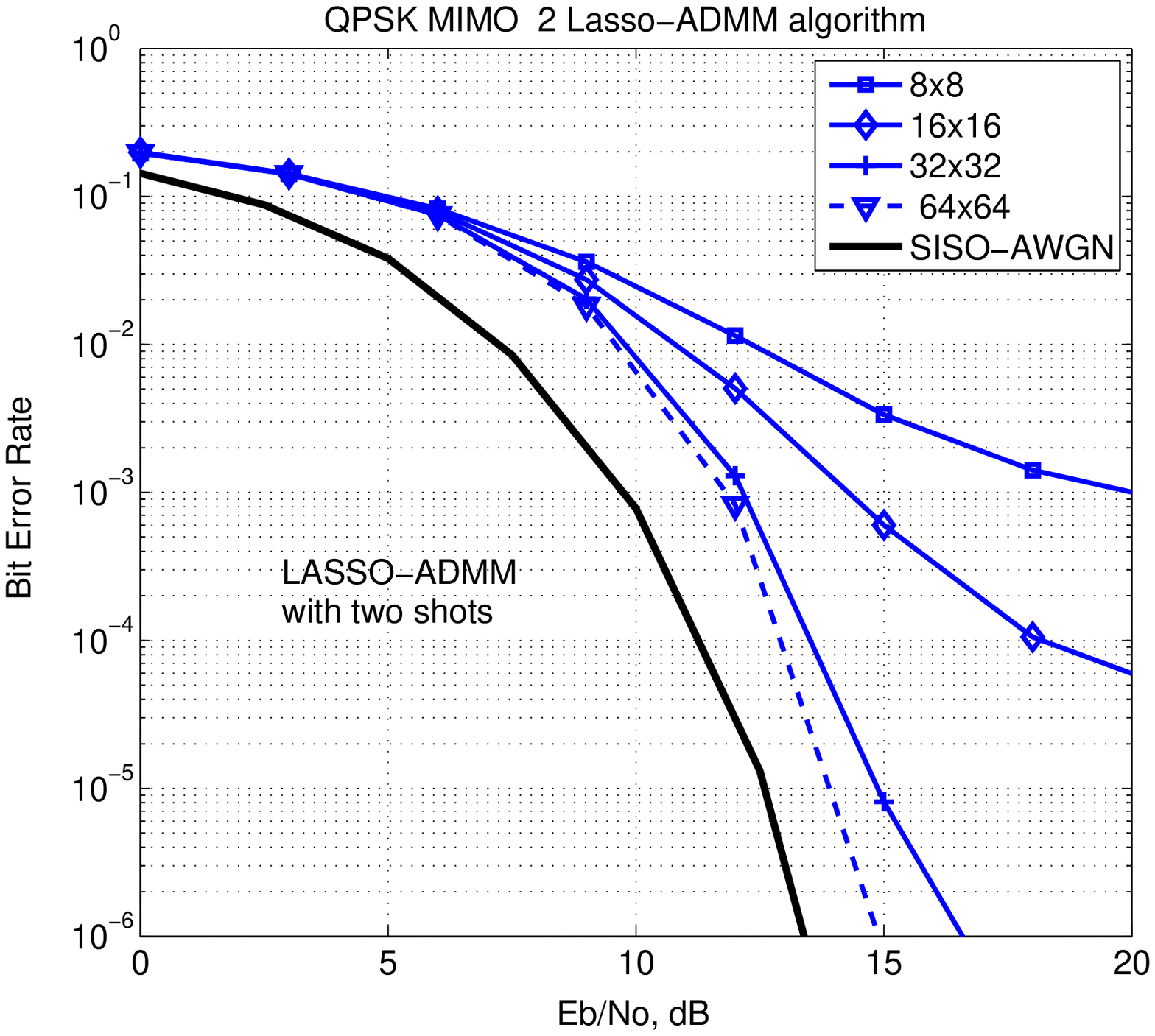}
%		\vspace{-.25in}
		\caption{\small QPSK BER performance of 2 Lasso ADMM Algorithm for various $N_t \times N_r$ MIMO }
		\label{fig: 2}
	\end{minipage}
%	\hspace{2mm}
	\begin{minipage}{.50\textwidth}
%				\centering
		\includegraphics[trim=0.0in 0.0in 4.50in 0.0in, clip,width=\textwidth]{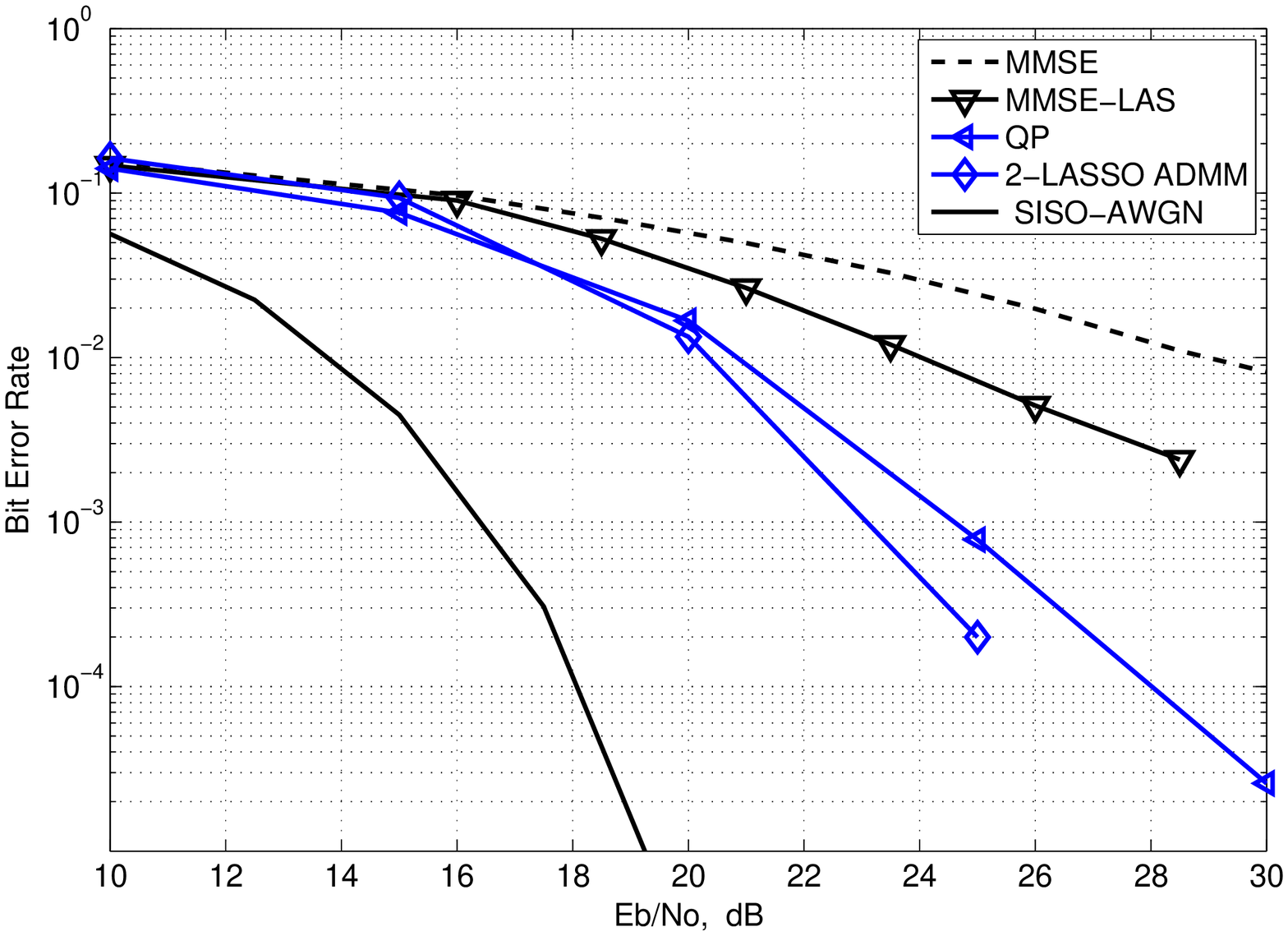}
%		\vspace{-.25in}
		\caption{\small BER performance of 16QAM $ 32\times 32$ MIMO}
		\label{fig:5}
	\end{minipage}
%\hspace{0.5mm}
	\begin{minipage}{.40\textwidth}
%		\centering
	\includegraphics[trim=0.1in 0.0in 0.20in 0.0in, clip,width=\textwidth]{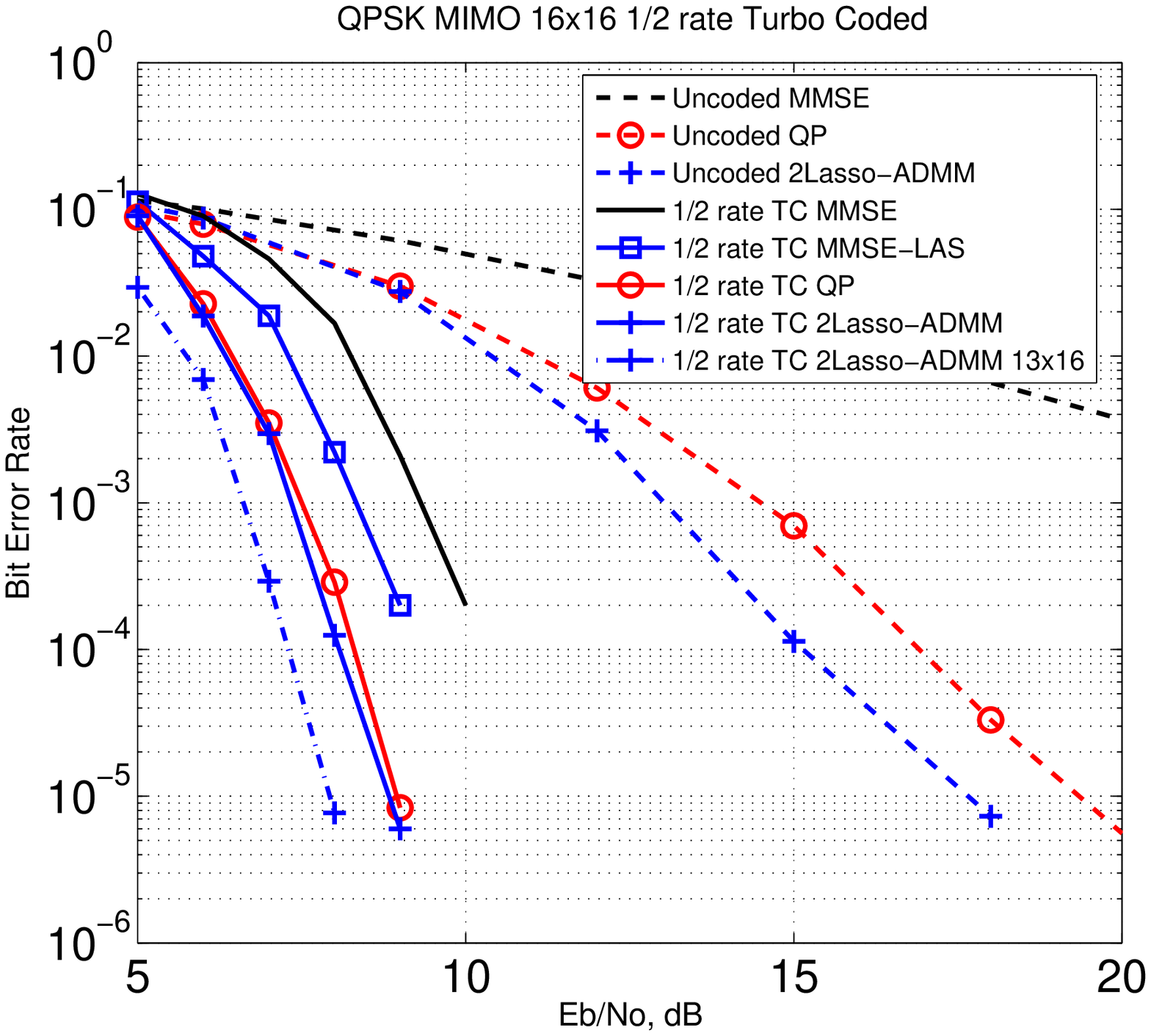}
	%\vspace{-.1in}
	\caption{\small $1/2$ Turbo coded BER performance of QPSK }
	%$16\times16$ 
	\label{fig: 7.0}
\end{minipage}
	\vspace{-0.25in}
%	\vspace{-.1in}
\end{figure*}

The proposed algorithms are referred to as LASSO-ADMM and 2-LASSO-ADMM depending on the number of shots the Algorithm is run. The focus of this paper is on the two-stage LASSO-ADMM algorithm. This algorithm is compared to some known and recent algorithms in large-scale MIMO, such as conventional minimum mean square estimator detector (MMSE), quadratic programming detector (QP), LAS, and RTS. For fair comparison between various detection techniques, all implementation is done using real system model.\\
Fig. \ref{fig: 1} shows clearly that there is an improvement of 2-LASSO-ADMM over LASSO-ADMM for $ 32\times 32$ MIMO configuration. It can also be seen that 2-LASSO-ADMM outperforms MMSE, LAS, QP, especially at SNR greater than 10 dB. At SNR less than 10 dB, all algorithms except MMSE performs more or less the same. At high SNR, our 2-LASSO-ADMM algorithm becomes  close to the AWGN single antenna limit with about 2 dB at BER= $10^{-5}$. The proposed algorithm is examined to see its performance when the  number of  antennas increases. This important aspect in large MIMO detectors referred to as adherence to a large system behavior. It means that the performance of a MIMO detector increases as the number of antennas increases \cite{elg015quad}. Fig. \ref{fig: 2} shows that the BER performance of the 2-LASSO-ADMM improves as $N_t \times N_r$ increases (e.g. , $ 8\times 8$, $ 16\times 16$, $ 32\times 32$, and $ 64\times 64$). For instance at BER= $10^{-5}$, the performance of $ 64\times 64$ is just about 1 dB away from SISO AWGN.
\indent 
%\begin{figure}[h]
%%%%\begin{minipage}[b]{1.0\linewidth}
%\centering
%\includegraphics[width=8cm, height=6cm]{./new_results/LASSO_2ADMM_qpsk_without_RTS_new_result.eps}
%%\includegraphics[width=8cm, height=6cm]{LASSO_2ADMM_qpsk_withRTS}
%\caption{\small BER performance of QPSK $ 32\times 32$ MIMO }
%\label{fig: 1}
%\end{figure}
%
%
%
%\begin{figure}[h]
%%%%\begin{minipage}[b]{1.0\linewidth}
%\centering
%\includegraphics[width=8cm, height=6cm]{LASSO_2ADMM_qpsk_various_NtxNr.eps}
%\caption{\small QPSK BER performance of 2 Lasso ADMM Algorithm for various $N_t \times N_r$ MIMO }
%\label{fig: 2}
%\end{figure}

\indent Fig. \ref{fig:5} shows that the 2-LASSO-ADMM algorithm performs well even at higher QAM modulations, such as 16QAM, especially at high SNR regime. It outperforms QP detector at SNR greater than 17 dB, and RTS at SNR greater than 23 dB. Although RTS performs slightly better than our algorithm at low SNR with 16QAM, it was shown at various references that RTS tends to have a degraded BER performance as SNR increases and also as the number of antennas increases, especially at higher QAM modulations \cite{elghariani2016low}.

% \cite{datta2011hybrid}, provides 2 dB improvement over the RTS.\\

%\begin{figure}[h]
%%%%\begin{minipage}[b]{1.0\linewidth}
%\centering
%\includegraphics[width=8cm, height=6cm]{./new_results/qam16_LASSO_ADMM_32x32_newresults.eps}
%%\includegraphics[width=8cm, height=6cm]{qam16_LASSO_ADMM_32x32_new.eps}
%\caption{\small BER performance of 16QAM $ 32\times 32$ MIMO}
%\label{fig:5}
%\end{figure}
%
%

%
% % % % % % % % % % % % % % % % % % % % % % % % % % % % % % % % % % % % % % % % %
%

%
\textit{{Turbo Coded BER Performance}} : The turbo coded BER performance of the Lasso ADMM detectors compared to MMSE, MMSE-LAS, and QP detectors is depicted in Fig. \ref{fig: 7.0} using QPSK modulation. In this simulation, $ 16\times 16 $ QPSK MIMO system is examined with rate-1/2 turbo encoder and decoder of 10 iterations. $ \pm1 $ output valued vector from all detectors is fed as an input to the BCJR-based turbo decoder. In Fig. \ref{fig: 7.0}, 2-LASSO-ADMM detector performs slightly better than QP, and clearly better than MMSE-LAS and MMSE detectors. The uncoded performance in this Figure is presented for reference only. It can be depicted in the last figure that by making $N_t < N_r$ with just 3 antennas, a significant improvement can be gained, which is about 1 dB at $10^{-4}$ BER. This configuration simulates the case of up link large multi-user MIMO, where base station has $N_r$ antennas and there are $N_t$ multi-user terminals each with single antenna.
%\begin{figure}[h]
%%%%\begin{minipage}[b]{1.0\linewidth}
%\centering
%\includegraphics[width=8cm, height=6cm]{QPSK_LASSO_ADMM_13x16_Turbo_encod.eps}
%\caption{\small 1/2 Turbo coded BER performance of QPSK $ 16\times16 $ }
%\label{fig: 7.0}
%\end{figure}
%
%

\vspace{-0.08in}

\section{Conclusion and Future work}

In this paper, a two-stage LASSO ADMM algorithm for large scale MIMO signal detection is investigated. Numerical results demonstrate the superiority of the proposed algorithm compared to the conventional methods for both coded and uncoded symbols at various QAM modulations. Investigating Multi-Stage LASSO-ADMM is an interesting future work. 
\vspace{-0.150in}
%We are currently investigating Multi-Stage LASSO ADMM Signal Detection Algorithm For Large Scale MIMO where $N$ stages are considered $(N > 2)$. Our preliminary results show further performance improvement when the rounding thresholds are well chosen per stage and for some values of $\lambda$ which interests us to understand in more depth the theory behind choosing these parameters..Finally,  we are working on Multi-Stage Generalized LASSO ADMM Signal Detection Algorithm for Spatial Modulation. In spatial modulation, not only the coefficients are sparse, but also the transmitted symbols since only one or few antennas is/are active at any given time.
\bibliographystyle{IEEEtran}

\bibliography{refMIMO}

\end{document}